\documentclass[12pt]{article}
\usepackage{graphicx}
\usepackage{indentfirst}
\usepackage{multirow}
\usepackage{amssymb}
\usepackage{amsfonts}
\usepackage{amsmath}
\usepackage{epsfig}
\usepackage{subfigure}

\def\beq{\begin{equation}}
\def\enq{\end{equation}}
\def\beqa{\begin{eqnarray}}
\def\enqa{\end{eqnarray}}

\def\qq{\lag\bar{q}q\rag}

\def\mix{\lag\bar{q}g\si.Gq\rag}

\def\gG{\lag g_s^2 G^2 \rag}
\def\G3{\lag g^3G^3\rag}
\def\pli{p^\prime}

\def\la{\lambda}

\def\ga{\gamma}
\def\Ga{\Gamma}

\def\si{\sigma}

\def\al{\alpha}

\def\lb{\label}
\def\nn{\nonumber}

\newcommand{\rag}{\rangle}
\newcommand{\lag}{\langle}

\newcommand{\rsup}[1]{\mbox{\tiny $#1$}}

\def\pbnr{}
\def\speaker{R.M. Albuquerque, J.M. Dias, M. Nielsen and C.M. Zanetti}
\def\onbehalfof{}
\def\title{$X(4260)$ as a Mixed Charmonium-Tetraquark State}
\def\affiliation{Institute for Theoretical Physics, S\~{a}o Paulo State University, Brazil (IFT/UNESP)\\
Institute of Physics, University of S\~ao Paulo, Brazil (IF/USP) \\
Faculty of Technology, Rio de Janeiro State University, Brazil (UERJ)}
\def\support{This work was supported by CNPq and FAPESP-Brazil.}

\newcommand\pubnumber{\pbnr}
\newcommand\pubdate{\today}
\def\Title#1{\begin{center} {\Large #1 } \end{center}}
\def\Author#1{\begin{center}{ \sc #1} \end{center}}

\newcommand{\OnBehalf}[1]{\sbox0{#1}\ifdim\wd0=0pt
        {}
	\else
	{\\on behalf of #1}
	\fi}
\newcommand{\SupportedBy}[1]{\sbox0{#1}\ifdim\wd0=0pt
        {}
	\else
	{\footnote{#1}}
	\fi}
\def\Address#1{\begin{center}{ \it #1} \end{center}}

\newcommand\pubblock{\includegraphics[width=5cm]{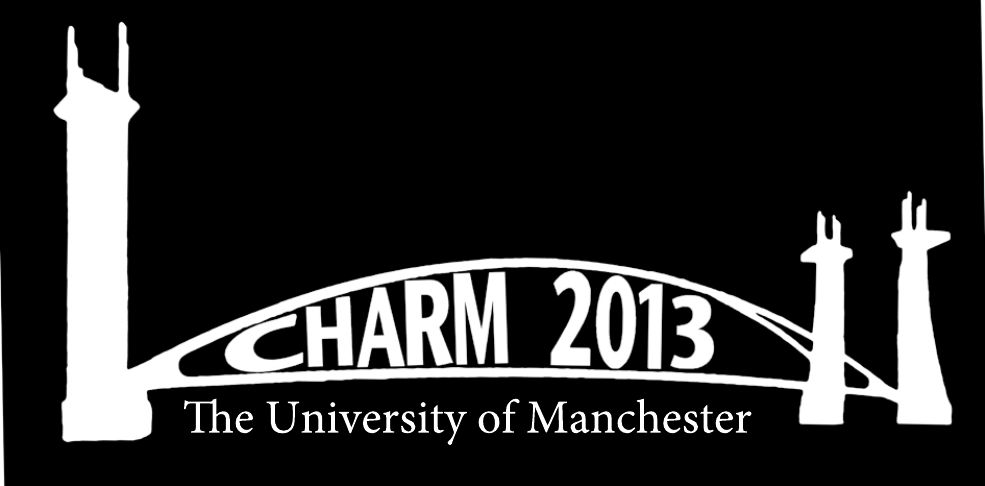}\hfill{\begin{tabular}{l} \pubnumber\\
         \pubdate  \end{tabular}}}
\newenvironment{Abstract}{\begin{quotation}  }{\end{quotation}}
\newenvironment{Presented}{\begin{quotation} \begin{center} 
             PRESENTED AT\end{center}\bigskip 
      \begin{center}\begin{large}}{\end{large}\end{center} \end{quotation}}

\def\venue{The 6$^{th}$ International Workshop on Charm Physics\\
(CHARM 2013)\\
Manchester, UK,  31 August -- 4 September, 2013}

\def\beq{\begin{equation}}
\def\eeq#1{\label{#1}\end{equation}}
\def\eeqn{\end{equation}}

\def\beqa{\begin{eqnarray}}
\def\eeqa#1{\label{#1}\end{eqnarray}}
\def\eeqan{\end{eqnarray}}

\let\bar=\overbar

\def\Dslash{\not{\hbox{\kern-4pt $D$}}}
\def\dslash{\not{\hbox{\kern-2pt $\del$}}}

\def\msb{{\bar{\ssstyle M \kern -1pt S}}}

\begin{document}
\begin{titlepage}
\pubblock

\vfill
\Title{\title}
\vfill
\Author{\speaker\SupportedBy{\support}\OnBehalf{\onbehalfof}}
\Address{\affiliation}
\vfill

\begin{Abstract}
Using the QCD sum rule approach we study the $X(4260)$ state assuming that 
it can be described by a mixed charmonium-tetraquark current with $J^{PC}= 1^{--}$ 
quantum numbers. For the mixing angle around $\theta \simeq (53.0 \pm 0.5)^o$, 
we obtain a value for the mass which is in good agreement with the experimental mass 
of the $X(4260)$. For the decay width into the channel $X \to J/\psi \pi\pi$ we find the 
value $\Gamma_{X \to J/\psi \pi\pi} \simeq (4.1 \pm 0.6) \:\mbox{MeV}$, which is much 
smaller than the total experimental width $\Gamma \simeq (108 \pm 12) \:\mbox{MeV}$. 
However, considering the experimental upper limits for the decay of the $X(4260)$ into 
open charm, we conclude that we cannot rule out the possibility of describing this state 
as a mixed charmonium-tetraquark state
\end{Abstract}
\vfill
\begin{Presented}
\venue
\end{Presented}
\vfill
\end{titlepage}
\def\thefootnote{\fnsymbol{footnote}}
\setcounter{footnote}{0}

\section{Introduction}
Recent results on charmonium spectroscopy carried out by {\sc Babar} and 
{\sc Belle} Collaborations revealed that many of the charmonium-like states 
observed in $e^+e^-$ collisions do not fit into the usual scheme quarkonia 
interpretation, and have stimulated an extensive discussion about exotic 
hadron configurations. 
Among these states, the $X(4260)$ was first observed by {\sc Babar} 
Collaboration in the $e^+e^-$ annihilation through initial state radiation 
\cite{babar1}, and it was confirmed by {\sc Cleo} and {\sc Belle} 
Collaborations \cite{yexp}. The $X(4260)$ was also observed in the 
$B^-\to X(4260)K^-\to J/\Psi\pi^+\pi^-K^-$ decay \cite{babary2}, and {\sc Cleo}
reported two additional decay channels: $J/\Psi\pi^0\pi^0$ and
$J/\Psi K^+K^-$ \cite{yexp}. 
One should notice that the $X(4260)$ mass is higher than the $D^{(*)}\bar{D}^{(*)}$
threshold, and if it was a normal $c\bar{c}$ charmonium state, it should decay 
mainly into this open-charm channel. However, this is not what was observed for 
this state \cite{belle5,babar5,babar6}.
Besides, the conventional $\Psi(3S),~\Psi(2D)$ and $\Psi(4S)$ $c\bar{c}$ states have 
been assigned to the well established $\Psi(4040),~\Psi(4160),~$ and 
$\Psi(4415)$ mesons, respectively, and the prediction from quark models 
for the $\Psi(3D)$ state is 4.52 GeV. Therefore, the $X(4260)$ mass is not consistent 
with any of the $1^{--}$ $c\bar{c}$ states \cite{Zhu:2007wz,Nielsen:2009uh,kz}. 
There are many theoretical interpretations for the $X(4260)$: tetraquark state 
\cite{tetraquark}, hadronic molecule of $D_{1} D$, $D_{0} D^*$ \cite{Ding}, 
$\chi_{c1} \omega$ \cite{Yuan}, $\chi_{c1} \rho$ \cite{liu}, $J/\psi f_0(980)$ \cite{oset}, 
a hybrid charmonium \cite{zhu}, a charm baryonium \cite{Qiao}, etc. 
Within the available experimental information, 
none of these suggestions can be completely ruled out. However, there are some
calculations, within the QCD sum rules (QCDSR) approach
\cite{Nielsen:2009uh,svz}, that can not explain 
the mass of the $X(4260)$ supposing it to be a tetraquark state \cite{rapha}, 
or  a $D_{1} D$, $D_{0} D^*$ hadronic molecule \cite{rapha}, or  a  
$J/\psi f_0(980)$ molecular state \cite{Albuquerque:2011ix}. 

In this work, we use again the QCDSR approach to evaluate both, mass and 
decay width, of the $X(4260)$ considering a new possibility for its structure: 
the mixing between two and four-quark states, which can be achieved with a 
mixed charmonium-tetraquark current in sum rules. For more details 
on this work please see the ref.\cite{prdY}.

\section{The Two- and Four-quark Operator}
In order to construct a mixed charmonium-tetraquark current, with 
$J^{PC}=1^{--}$, we have to define the currents associated with the 
charmonium and the tetraquark states. For the charmonium part, we 
use the conventional charmonium vector current: $j_\mu^{'(2)}=\bar{c}_a \ga_\mu c_a$,
while the tetraquark part is interpolated by \cite{rapha}
\beqa
j_\mu^{(4)} &=& \frac{\epsilon_{abc} \epsilon_{dec}}{\sqrt{2}}
\Big[ (q_a^T C\ga_5 c_b)(\bar{q}_d \ga_\mu\ga_5 C\bar{c}_e^T) +
(q_a^T C\ga_5\ga_\mu c_b)(\bar{q}_d \ga_5 C\bar{c}_e^T) \Big].
\label{j4q}
\enqa
As in Refs. \cite{oka24,matheus}, we define the normalized two-quark
current as 
\beq
j_\mu^{(2)}=\frac{1}{\sqrt{2}}\qq ~j_\mu^{'(2)}.
\enq
Then using these two currents we build the following mixed
charmonium-tetraquark current for the $X(4260)$ state:
\beq
j_\mu(x)=\sin(\theta) \:j_\mu^{(4)}(x)+\cos(\theta) \:j_\mu^{(2)}(x) ~~.
\label{jmix}
\enq

\section{The Two-Point Correlation Function}
To calculate the mass of a hadronic state using the QCDSR approach, 
the starting point is the two-point correlation function
\begin{align}
\Pi_{\mu\nu}(q) &= i \!\!\int \!d^{4}x ~e^{i q\cdot x}
\langle 0| \,T[ j_{\mu}(x) j_{\nu}^{\dagger} (0) ] \,|0 \rangle =
 -\Pi_{1}(q^{2})\Big(g_{\mu \nu} - \frac{q_{\mu}q_{\nu}}{q^{2}}
 \Big) + \Pi_{0}(q^{2})
 \frac{q_{\mu}q_{\nu}}{q^{2}},
\label{2point}
\end{align}
where $j_{\mu}(x)$ is given by Eq.~(\ref{jmix}). The functions 
$\Pi_{1}(q^{2})$ and $\Pi_{0}(q^{2})$ are two independent invariant 
functions related to spin-1 and spin-0 mesons, respectively.
The two-point correlation function can be evaluated in two ways,
according to the principle of duality: in the OPE side, we calculate 
it in terms of quarks and gluon fields using the Wilson's operator 
product expansion (OPE). In the phenomenological side, we insert a 
complete set of intermediate states with $1^{--}$ quantum numbers, 
and we parametrize the coupling of the vector state $X$ with the current, 
defined in Eq. (\ref{jmix}), through the coupling parametrization: 
$\langle 0| j_{\mu}(x)|X\rangle = \lambda_X \epsilon_{\mu}$
where $\epsilon_\mu$ is the polarization vector. Thus, we can write the 
phenomenological side of Eq. (\ref{2point}) as
\begin{equation}
\Pi^{\rsup{PHEN}}_{\mu \nu}(q) = \frac{\lambda_X^{2}}{M_{X}^{2} - q^{2}}
\Big(g_{\mu \nu} - \frac{q_{\mu}q_{\nu}} {q^{2}} \Big) + \:.\:.\:.\:
\label{phenoside}
\end{equation}
where $M_X$ is the mass of the $X$ state and the dots represent the 
higher resonance contributions which will be parametrized, as usual, 
through introduction of the continuum threshold parameter $s_{0}$ 
\cite{io1}. 
The OPE side can be written in terms of a dispersion relation
\begin{equation}
\Pi^{\rsup{OPE}}(q^{2}) = \int\limits_{4m^{2}_c}^{\infty}
 ds\frac{\rho^{\rsup{OPE}}(s)}{s - q^{2}},
 \label{opeside}
\end{equation}
where $\rho^{\rsup{OPE}}(s)$ is the spectral density and can be obtained by:
$\pi \rho^{\rsup{OPE}}(s) = \mbox{Im}[\Pi^{\rsup{OPE}}(s)]$. 
In this side, we work at leading order in $\alpha_{s}$ in the operators and 
we consider the contributions from the condensates up to dimension-8 in 
the OPE. After making a Borel transform in the equations (\ref{phenoside}) and 
(\ref{opeside}), we are able to match both sides of the correlation function in 
order to extract the mass of the charmonium-tetraquark state.

\subsection{Numerical Analysis}
\begin{figure}[tp]
\begin{center}
\centerline{\includegraphics[width=0.48\textwidth]{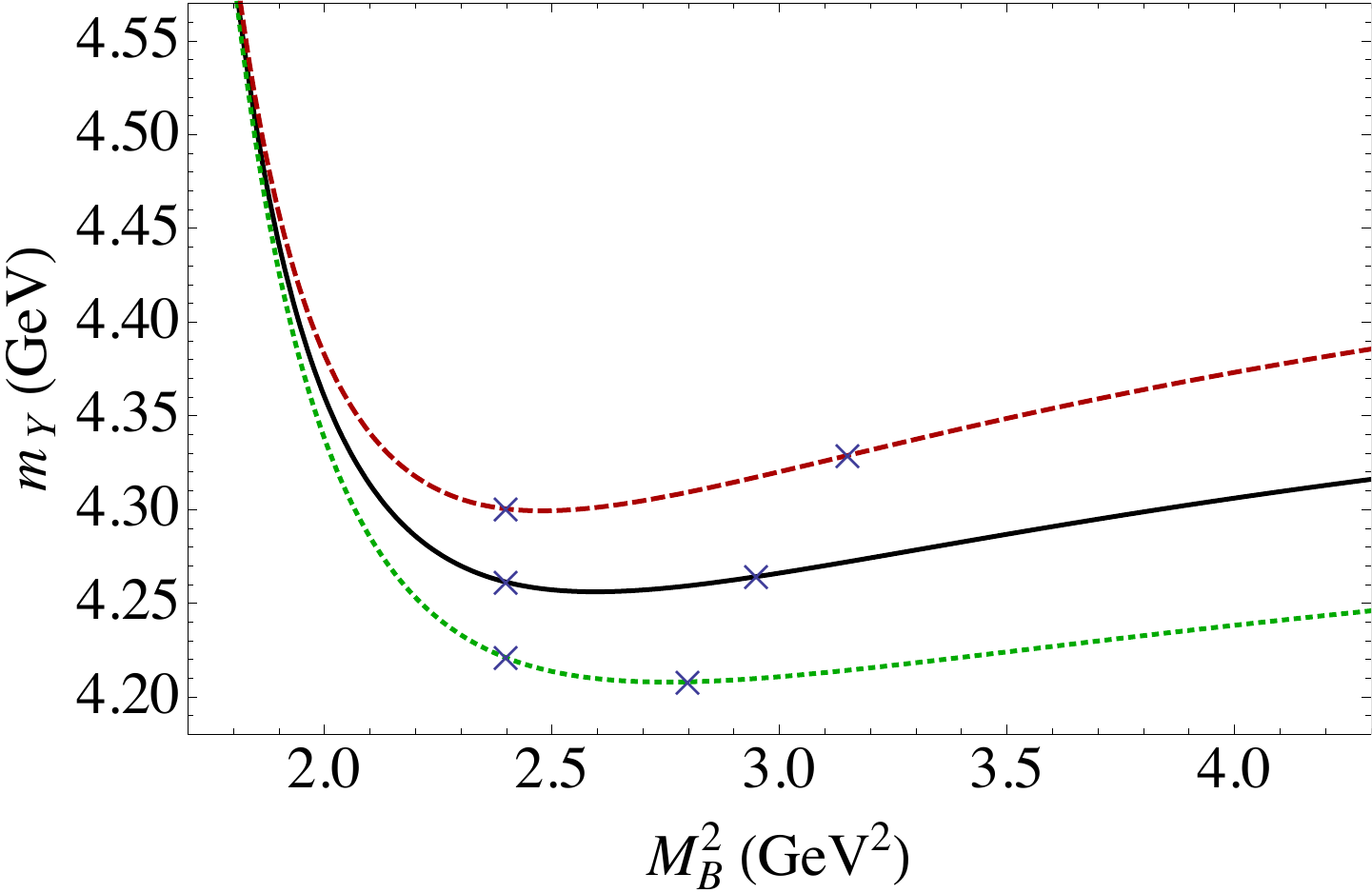}}
\caption{The mass as a function of the sum rule parameter 
$M^{2}_{B}$ for $\sqrt{s_{0}} = 4.60$ GeV (dotted line), 
$\sqrt{s_{0}} = 4.70$ GeV (solid line), $\sqrt{s_{0}} = 4.80$ GeV
(long-dashed line). The crosses indicate the valid Borel Window.}
\label{fig3} 
\end{center}
\vspace{-1.0cm}
\end{figure}

In Table \ref{Param}, we list the numerical values of the quark 
masses and condensates that we have used in our sum rule analysis. 
{\small
\begin{table}[t]
\begin{center}
\setlength{\tabcolsep}{1.25pc}
\caption{Quark masses and condensates values 
\cite{Albuquerque:2011ix,x3872,narpdg}.}
\begin{tabular}{ll}
&\\
\hline
Parameters&Values\\
\hline
$m_{c}(m_{c})$ & $(1.23 \pm 0.05) ~\mbox{GeV}$ \\
$\langle \bar{q}q \rangle$ & $-(0.23 \pm 0.03)^{3} ~\mbox{GeV}^{3}$\\
$\langle \bar{q}g\sigma.Gq \rangle$ & $m_{0}^{2}\langle \bar{q}q\rangle$\\
$m_{0}^{2} $ & $(0.8 \pm 0.1)~\mbox{GeV}^{2}$\\
$\langle g_{s}^{2} G^{2} \rangle$ & $(0.88\pm 0.25)~\mbox{GeV}^{4}$\\
\hline
\end{tabular}
\label{Param}
\end{center}
\vspace{-0.5cm}
\end{table}}
The continuum threshold, $\sqrt{s_0}$, is a physical parameter that  
should be related to the first excited state with the same quantum numbers. 
Since the spectrum of the mixed state, given by Eq.(\ref{jmix}), is completely 
unknown we will fix the continuum threshold range starting with the smaller 
value which provides a valid Borel window. Using this criterion, we obtain 
$s_{0}$ in the range $4.6 \leq \sqrt{s_{0}} \leq 4.8$ GeV.
Notice that reliable results from the sum rule approach only can be obtained 
establishing a valid Borel Window. This condition is satisfied imposing a
good OPE convergence, the pole dominance over the continuum contribution 
and a good Borel stability. Then after we have determined the Borel window, 
we can calculate the ground state mass, which is shown, as a 
function of $M^{2}_{B}$, in the Fig. \ref{fig3}.

We can reproduce the experimental mass of the $X(4260)$, 
$M_X = 4250 \pm 9$ MeV, setting the value of the mixing angle as
\begin{eqnarray}
\theta=(53.0\pm0.5)^0 ~,  
\end{eqnarray}
altogether with the variations of other parameters as 
indicated in Table I, and considering the continuum threshold in the range 
$\sqrt{s_{0}} = 4.70 \pm 0.10$ GeV.
Thus, we can also estimate the meson-current coupling parameter.
Using the same values of the $s_0$, $\theta$ and the Borel Window 
used for the mass calculation, we get:
\beq
\lambda_X = (2.00 \pm 0.23) \times 10^{-2} ~ \mbox{GeV}^5.
\label{lay}
\enq

\section{The $X(4260)$ Decay Modes}
The QCDSR technique can also be used to evaluate the
coupling constants and form factors for a given vertex. Indeed, 
the authors in Ref.\cite{bcnn} determined the form factors and coupling 
constants for many hadronic vertices containing charmed mesons, by 
using the QCD sum rules method.

First, we evaluate the coupling constant associated with the vertex 
$X \:J/\psi \:\si$ to estimate the decay width of the process 
$X \rightarrow J/\psi \:\pi \pi$. We assume that the two pions in the 
final state come from the $\sigma$ meson. 
In order to determine this coupling constant, we must calculate the 
three-point function defined as

\beq
\Pi_{\mu \nu}(p,p^\prime, q) = \int \!d^4x \:d^4y ~e^{i p^\prime \cdot x}
  \:e^{iq\cdot y} ~\lag 0|T\{j_{\mu}^{\psi}(x)j^{\sigma}(y)j_{\nu}^{X \:\dagger}(0)\}|0\rag 
\label{3po}
\enq
with $p= p^\prime+q$.
The respective interpolating fields are given by the currents of $J/\psi$, 
$\sigma$ and $X(4260)$ states. For the $\sigma$ meson current we use:
$j^{\sigma}=\frac{1}{\sqrt{2}}\Big(\bar{u}_a u_a
+ \bar{d}_a d_a\Big)$.

The three-point correlation function can also be described in terms 
of hadronic degrees of freedom (Phenomenological side) as well as 
in terms of quarks and gluons fields (OPE side). In order 
to evaluate the phenomenological side of the sum rule we  
insert, in Eq.(\ref{3po}), intermediate states for $X$, 
$J/\psi$ and $\sigma$. Using the definitions: 
\vspace{-0.2cm}
\begin{eqnarray}
\lag 0 | j_\mu^\psi|J/\psi(\pli)\rag = M_\psi 
f_{\psi}\epsilon_\mu(\pli), ~~~~~~
\lag 0 | j^\sigma|\sigma(q)\rag = A_{\si}, ~~~~~~
\lag X(p) | j_\nu^X|0\rag = \la_X \epsilon_\nu^*(p), \nn
\enqa
we obtain the following relation:
\vspace{-0.2cm}
\beqa
\Pi_{\mu\nu}^{\rsup{PHEN}} (p,\pli,q) &\!\!\!=\!\!\!&
{\la_X M_{\psi} f_{\psi} A_\sigma \:g_{_{X \psi\sigma}}(q^2)
\over (p^2 \!-\! M_{X}^2)({\pli}^2 \!-\! M_{\psi}^2) (q^2 \!-\! M_\sigma^2)}
\Big[ (\pli \cdot p) g_{\mu\nu} \!-\! \pli_\nu q_\mu \!-\!  \pli_\nu \pli_\mu \Big] +\cdots \:, ~~~~~
\lb{phen}
\enqa
where the dots stand for the contribution of all possible excited states. 
The form factor, $g_{_{X\psi \si}}(q^2)$, is defined by the generalization 
of the on-shell mass matrix element, $\lag J/\psi \sigma|X\rag$, for an 
off-shell $\sigma$ meson: 
\vspace{-0.3cm}
\beq
\lag J/\psi\si |X\rag= g_{_{X\psi \sigma}}(q^2)
\Big[ \pli \cdot p ~\epsilon^*(\pli)\cdot \epsilon(p) -
\pli \cdot \epsilon(p)~p\cdot \epsilon^*(\pli) \Big],
\label{coup}
\enq
which can be extracted from the effective Lagrangian 
that describes the coupling between two vector mesons 
and one scalar meson:
${\cal{L}}=ig_{_{X \psi \sigma}}V_{\alpha\beta} A^{\alpha \beta}~\sigma$, 
where $V_{\alpha \beta} = \partial_{\alpha} X_{\beta} - \partial_{\beta} X_{\alpha}$ 
and $A^{\alpha \beta} = \partial^{\alpha} \psi^{\beta} - \partial^{\beta} \psi^{\alpha}$, 
are the tensor fields of the $X$ and $\psi$ fields respectively.
In the OPE side, we work at leading order in $\al_s$ and
we consider the condensates up to dimension-5. 
Taking the limit $p^2 = {\pli}^2=-P^2$ and doing the Borel transform 
to $P^2 \rightarrow M^2$, we get the following expression in the structure ${\pli}_{\nu} q_\mu$:
\beqa
\frac{\lambda_X A_{\si} M_\psi f_\psi \:g_{_{X\psi \sigma}}(Q^2)}
{(M_X^2 \!-\! M_\psi^2)(Q^2 \!+\! M_\sigma^2)} \left(e^{-M_\psi^2/M^2}
\!- e^{-M_X^2/M^2}\right) \!+\! B(Q^2)~e^{-s_0/M^2} = 
\Pi^{\rsup{OPE}}(M^2,Q^2),
\label{3sr}
\enqa
where $Q^2=-q^2$, and $B(Q^2)$ gives the contribution to 
the pole-continuum transitions \cite{matheus,io2,decayx,dsdpi}.
The $M_{\psi}$ and $f_{\psi}$ are the mass and decay constant of 
the $J/\psi$ meson and $M_{\si}$ is the mass of the $\si$ meson. Their values are 
given by: $M_{\psi} = 3.1$ GeV, $f_{\psi}=0.405$ GeV \cite{pdg}, and 
$M_{\si}=0.478$ GeV \cite{ignacio}. The parameters $\la_{X}$ and 
$A_{\si}$ represent the couplings of the $X$ and $\si$ states with the respective 
currents. The value of $\la_{X}$ is given by Eq.\,(\ref{lay}), while $A_{\si}$ was 
calculated in Ref. \cite{dosch} and its numerical value is $A_{\si}=0.197$ GeV$^2$.
Finally, the $\Pi^{\rsup{OPE}}(M^2,Q^2)$ function is given by
\beqa
\Pi^{\rsup{OPE}}(M^2,Q^2) &\!\!\!\!=\!\!\!\!& \frac{\mbox{sin}(\theta)}{48\sqrt{2}\:\pi^2}
\!\int\limits_{0}^{1} \!\!d\al \,e^{-\frac{m_c^2/M^2}{\al (1\!-\!\al)}}\!
\bigg[ \frac{m_c\mix}{Q^2}
\bigg( \!\frac{1 \!-\! 2\al(1 \!-\! \al)}{\al(1\!-\!\al)} \!\bigg) \!-\! \frac{\gG}{2^5\pi^4}\bigg] .
~~~~~~
\label{opeside}
\enqa
%
%
\begin{figure}[t]
\centering
\subfigure[]{
\includegraphics[width=0.47\textwidth]{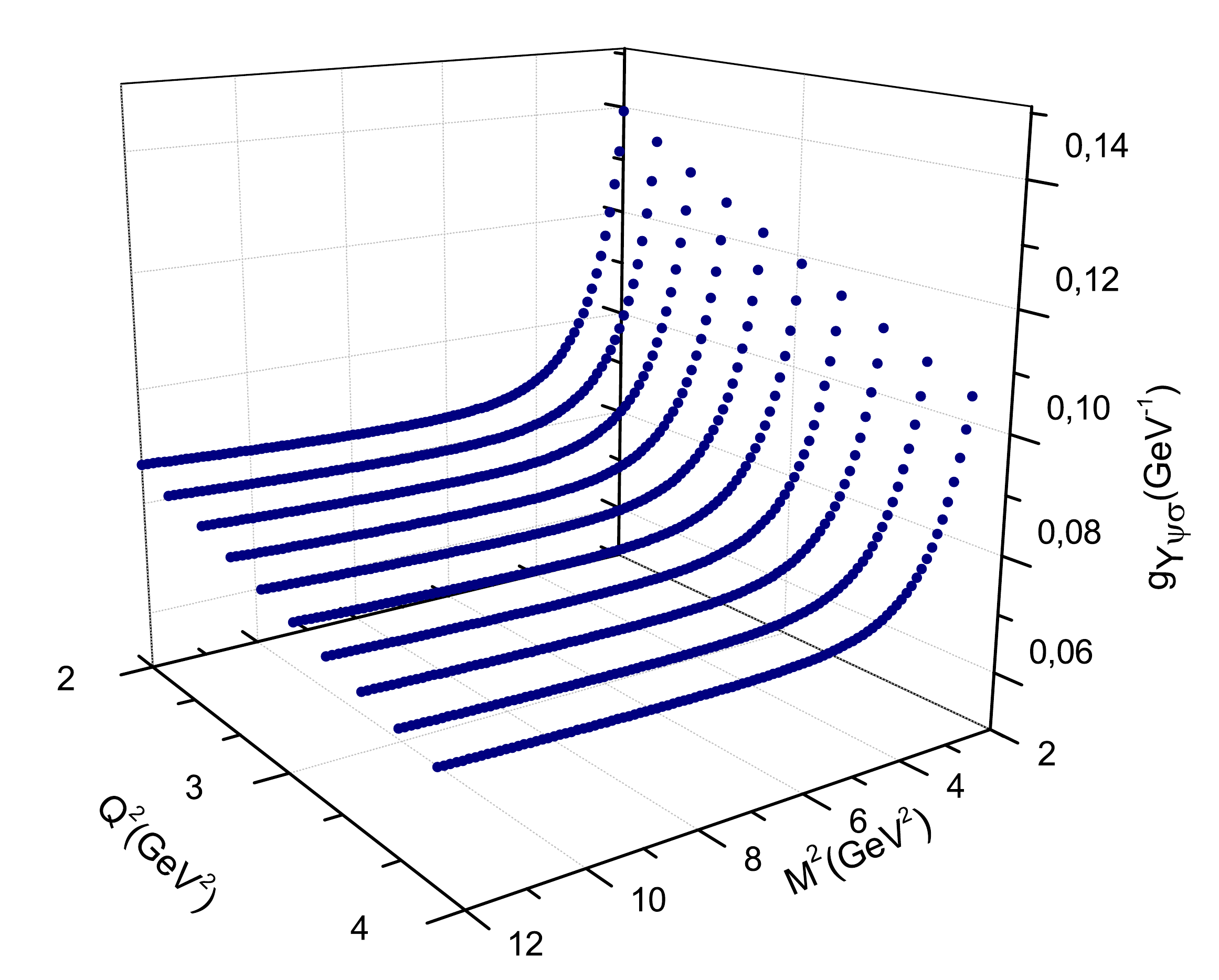}
}
\subfigure[]{
\includegraphics[width=0.47\textwidth]{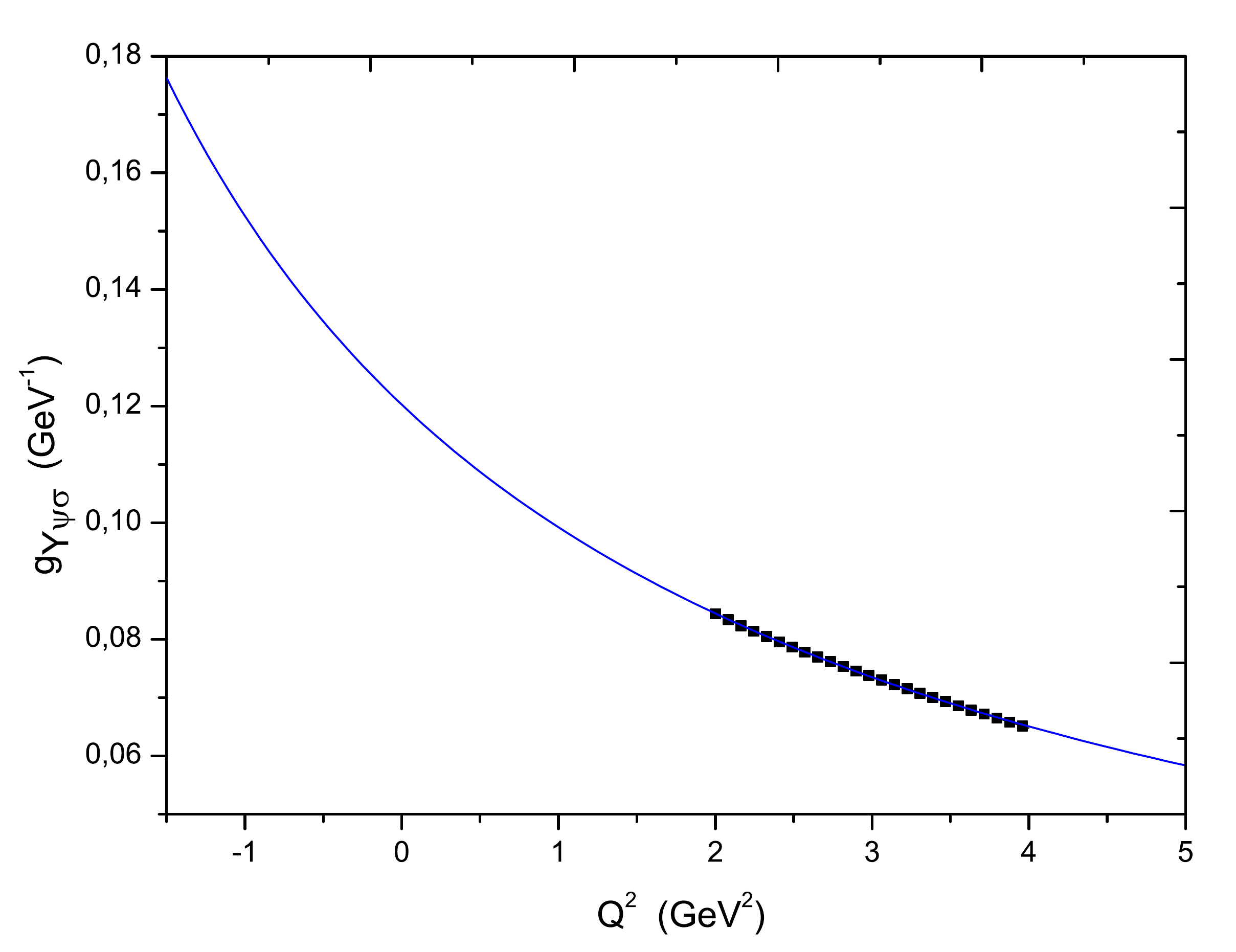}
}
\caption{{\bf a)} $g_{_{X\psi\si}}(Q^2)$ values obtained by varying both $Q^2$ and $M^2$.
{\bf b)} QCDSR results for $g_{_{X\psi\si}}(Q^2)$, as a function of $Q^2$, 
for $\sqrt{s_0}=4.76$ GeV (squares). 
The solid line gives the parametrization of the QCDSR results 
 through Eq. (\ref{mono}).}
\label{fig3d}
\end{figure}
In the sum rule of the three-point correlator we are interested in determine a 
region in the Borel mass where the form factor is independent of $M^2$.
In Fig.\,\ref{fig3d}a), we plot the $g_{_{X\psi\si}}(Q^2)$ as a function of
both $M^2$ and $Q^2$. Notice that in the region 
$7.0 \leq M^2 \leq 10.0$ GeV$^2$, the form factor is stable, as a function 
of $M^2$, for all values of $Q^2$.
In Fig.\,\ref{fig3d}b), we plot the $Q^2$ dependence 
of $g_{_{X\psi\si}}(Q^2)$, obtained for $M^2=8.0$ GeV$^2$.
Therefore, in order to calculate the coupling constant, we must estimate the 
value of the form factor at the meson pole: $Q^2 = -M^2_{\si}$. 
For this purpose, we need to extrapolate the form factor to the region of $Q^2$ 
where the sum rule method is not applicable. Such extrapolation can be done 
by parametrizing the $g_{_{X\psi\si}}(Q^2)$ form factor using a monopole form:
\vspace{-0.1cm}
\beq
	g_{_{X\psi\si}}(Q^2) = \frac{g_1}{g_2 + Q^2}.
	\label{mono}
\enq
Using this monopolar fit to the data indicated by the squares in Fig.\,\ref{fig3d}b), 
we found the following parameters:
$g_1 = (0.58~\pm ~0.04)~\mbox{GeV}$ and $g_2=(4.71~\pm ~0.06)~\mbox{GeV}^2$.  
The solid line in Fig.\,\ref{fig3d}b) shows that the parametrization given by 
Eq.\,(\ref{mono}) fits quite well the data for $g_{_{X\psi\si}}(Q^2)$.
Finally, the coupling constant $g_{_{X\psi\si}}$ is given by:
\vspace{-0.2cm}
\beq
g_{_{X\psi\sigma}}=g_{_{X\psi\sigma}}(-M^2_\sigma)=(0.13 \pm 0.01)~~\mbox{GeV}^{-1}.
\label{coupvalue}
\enq
The main source of uncertainty comes from the variations of $s_0$ and $\theta$.
The decay width for the process $X(4260) \rightarrow J/\psi \sigma \rightarrow J/\psi\pi \pi$
in the narrow width approximation is given by
\vspace{-0.2cm}
\beqa
{d\over ds}\Gamma_{_{X\to J/\psi \:\pi\pi}} &\!\!\!=\!\!\!& \frac{|{\cal{M}}|^2}{8\pi M_X^2}
\left(\frac{M_X^2 -M^2_{\psi}+s}{2M_X^2}\right)
{\Ga_{\si}(s) M_{\si}\over\pi}{p(s)\over(s-M_{\si}^2)^2+
(M_{\si}\Ga_{\si}(s))^2},
\label{de1}
\enqa
with $p(s)$ given by
$p(s)={\sqrt{\la(M_X^2,M_\psi^2,s)}\over2M_X}$, 
where $\la(a,b,c)=a^2+b^2+c^2-2ab-2ac-2bc$, and $\Ga_{\si}(s)$ 
is the s-dependent width of an off-shell $\sigma$ meson \cite{ignacio}:
\vspace{-0.2cm}
\beq
\Ga_{\si}(s)=\Ga_{0\si}\sqrt{\frac{\la(s,M_{\pi}^2, M_{\pi}^2)}
{\la(M_X^2,M_{\pi}^2, M_{\pi}^2)}}\frac{M_X^2}{s},
\enq
where $\Ga_{0\si}$ is the experimental value for the decay 
of the $\si$ meson into two pions. Its value is 
$\Ga_{0\si}=(0.324\pm 0.042\pm 0.021)$ GeV \cite{ignacio}.
The invariant amplitude squared can be obtained from 
the matrix element in Eq. (\ref{coup}). We get:
\beq
|{\cal M}|^2= \frac{g_{_{X\psi \sigma}}^2(s)}{3} 
\left[ M_X^2 M_\psi^2 + {1\over2}(M_X^2 + M_\psi^2 - s)^2\right]
\enq
Therefore, the decay width for the process 
$X(4260)\rightarrow J/\psi \:\pi\pi$ is given by
\beq
\Ga_{_{X\to J/\psi \:\pi\pi}} = \frac{M_\si}{16\pi^2 M^4_X} \!\!\!\!\int\limits_{4M_\pi^2}^{(M_X - M_\psi)^2} \!\!\!\!\!\!ds 
\:|{\cal M}|^2 \:\Ga_\si(s) \:p(s)
\frac{(M^2_X-M^2_\psi +s) }{(s-M^2_\si)^2+(M_\si \Ga_\si(s))^2}.
\label{width}
\enq
Hence, taking variations on $s_0$ and $\theta$ in the same intervals 
given above, we obtain from Eqs. (\ref{coupvalue})-(\ref{width}) the 
following value for the decay width
\beq
\Ga^\si_{_{X\to J/\psi \:\pi\pi}} = (1.0\pm 0.4) ~\mbox{MeV} ~~.
\enq
Doing the same analysis presented before, but now considering some 
adjustments \cite{prdY} for each channel, we can proceed 
to estimate the decay widths related to other processes like
$X \to J/\psi \:f_0(980) \to J/\psi \:\pi\pi$ and $X \to  J/\psi \:f_0(980) \to J/\psi \:KK$. 
Hence, we find \cite{prdY} 
\vspace{-0.3cm}
\beqa
\Ga^{f_0}_{_{X\to J/\psi \:\pi\pi}} &=& (3.1\pm 0.2) ~\mbox{MeV} ~~, \\
\Ga^{f_0}_{_{X\to J/\psi \:KK}} &=& (1.3\pm 0.4) ~\mbox{MeV} ~~.
\enqa
Our estimation for the total width is given by $\Ga_{_{tot}} \simeq 5.4 \pm 1.0$ MeV,
which is much smaller than the experimental data $\Ga_{_{exp}} = 108 \pm 12$ MeV.

\section{Summary and Conclusions}
In summary, we have used the QCDSR approach to study the two-point and 
three-point functions of the $X(4260)$ state, by considering a 
mixed charmonium-tetraquark current. A very good agreement with the
experimental value of the $X(4260)$ mass is achieved for the mixing angle 
around $\theta \simeq (53.0\pm 0.5)^{0}$. 
To evaluate the width of the decay $X(4260)\to J/\psi\pi\pi$, we work 
with the three-point function. First, we assume that the two pions in the final 
state come from the $\sigma$ and $f_0(9800$ scalar mesons. We also consider 
the process $X(4260)\to J/\psi \:KK$ with the $f_0(980)$ as an intermediate state. 
The obtained value for width is $\Ga_X \simeq (5.4\pm 1.0)$ MeV, which is much 
smaller than the experimental data: $\Ga_{exp} \simeq (108 \pm 12)$ MeV.
Possibly the main decay channel of the $X(4260)$ should be into $D$ mesons, 
mostly due to the presence of charmonium in its internal structure. These channels 
could increase the value estimated for $\Ga_X$.
Therefore, our findings indicate that an exotic hadronic structure for the $X(4260)$
cannot be ruled out. Indeed, a mixed charmonium-tetraquark state is a good candidate 
for explaining the mass and the decay channels observed experimentally.

\end{document}